# Quantifying Cortical Bone Free Water Using short echo time (STE-MRI) at 1.5T

**Shahrokh Abbasi-Rad**[a,b], **Atena Akbari**[a,b], **Malakeh Malekzadeh**[a,b,c*], **Mohammad Shahgholi**[a,d,e],

**Hossein Arabalibeik**[f], **Hamidreza Saligheh Rad**[a,b,g,*]

[a]Quantitative Medical Imaging Systems Group (QMISG), Research Center for Molecular and Cellular Imaging (RCMCI), Tehran Unievrsity of Medical Sciences (TUMS), Tehran, Iran.
[b]Department of Medical Physics and Biomedical Engineering, Tehran University of Medical Sciences (TUMS), Tehran, Iran.
[c]Medical Physics Department, School of Medicine, Iran University of Medical Sciences (IUMS), Tehran, Iran.
[d]Department of Mechanical Engineering, Najafabad Branch, Islamic Azad University, Najafabad, Iran.
[e]Modern Manufacturing Technologies Research Center, Najafabad Branch, Islamic Azad University, Najafabad, Iran.
[f]Research Center of Biomedical Technology and Robotics, Tehran University of Medical Sciences (TUMS), Tehran, Iran.
[g]Osteoporosis Research Center, Endocrinology and Metabolism Clinical Sciences Institute, Tehran University of Medical Sciences, Tehran, Iran.

**Corresponding Author:**
- **Shahrokh Abbasi-Rad**, Research Center for Molecular and Cellular Imaging, Institute for Advanced Medical Technologies, Imam Hospital, Keshavarz Blvd, 1419733141 Tehran, Iran. Email: Shahrokh.zita@gmail.com

* Author MM and author HSR contributed equally to this work.

# Abstract

**Purpose**: The purpose of our study was to use Dual-TR STE-MR protocol as a clinical tool for cortical bone free water quantification at 1.5T and validate it by comparing the obtained results (MR-derived results) with dehydration results.

**Methods**: Human studies were compliant with HIPPA and were approved by the institutional review board. Short Echo Time (STE) MR imaging with different Repetition Times (TRs) was used for quantification of cortical bone free water $T_1$ ($T_1$free) and concentration ($\rho_{free}$). The proposed strategy was compared with the dehydration technique in seven bovine cortical bone samples. The agreement between the two methods was quantified by using Bland and Altman analysis. Then we applied the technique on a cross-sectional population of thirty healthy volunteers (18F/12M) and examined the association of the biomarkers with age.

**Results**: The mean values of $\rho_{free}$ for bovine cortical bone specimens were quantified as 4.37% and 5.34% by using STE-MR and dehydration techniques, respectively. The Bland and Altman analysis showed good agreement between the two methods along with the suggestion of 0.99% bias between them. Strong correlations were also reported between $\rho_{free}$ ($r^2 = 0.62$) and $T_1$free and age ($r^2 = 0.8$). The reproducibility of the method, evaluated in eight subjects, yielded an intra-class correlation of 0.95.

**Conclusion**: STE-MR imaging with dual-TR strategy is a clinical solution for quantifying cortical bone $\rho_{free}$ and $T_1$free.

## 1. Introduction

Bone quality has been intriguing the scientists since many years ago and different researchers have introduced many techniques and methods for bone quality assessment. These techniques are majorly based on X-Ray imaging, quantifying Bone Mineral Density (BMD), which is considered to yield bone properties including stiffness, toughness, and strain (1-4). Given the fact that BMD only accounts for 30-50 % of fractures (5), and considering that cortical bone consists approximately 20-25% water by volume, it can be concluded that bone quality assessment should not be limited to only BMD measurement (6).

Recent investigations on *cortical bone water* have shown that it has rich and conclusive information which is shifting scientists' attention towards new insights into the bone quality assessment (7, 8). Water molecules occur at three different locations in cortical bone. A large fraction is either covalently bonded to the crystals of the apatite-like minerals (tightly bound) or participating in the hydrogen bonding with hydrophilic side chains of the proteins in the organic matrix of collagen (loosely bound), which is called 'bound water.' It is representative of bone flexibility or its resistance to fracture. A smaller fraction resides freely in the pores of cortical bone (Haversian canals, lacuna, and the canalicular system) which is called 'free' or 'pore' water, which characterizes cortical bone fragility (7, 9, 10).

Free water can provide a surrogate measure of porosity, which is hard to be measured directly by *in vivo* imaging modalities. In addition, during aging and osteoporosis, an increase in free volume fraction takes place, which consequently results in an increase in free water (11). Therefore, one can hypothesize that cortical bone free water might contribute to model the age-related increase of cortical bone porosity. This highlights the need to develop a method for quantifying cortical bone free water.

Magnetic Resonance Imaging (MRI) due to its sensitivity to proton micro-environments, is a good candidate for discovering cortical bone water, where Ultrashort Echo time (UTE) pulse sequence has been a pillar of the quantification (12-15). UTE was used with bi or tri-component exponential fitting of $T_2^*$ weighted signal to separately quantify different proton pools (16-21). In another study, a genetic algorithm was used as the optimization technique for solving a model-based inverse problem using the UTE signal to quantify free and bound water (22). Another group of UTE-MRI techniques used the advantage of adiabatic pulses to selectively image free and bound water using Dual Adiabatic Full Passage (DAFP) and Adiabatic Inversion Recovery (AIR) preparations for UTE imaging, respectively (23-27). In a recent study, magnetization transfer imaging was combined with UTE to differentiate between three proton pools named as bound, free and macromolecular proton fractions (28).

Although efforts made to adopt the UTE sequences so that they could widely be used in a clinical setting (29), but the sequence itself is not yet readily available in most clinical environments. In the previous study, short echo time (STE-MRI) was used as a clinical solution to quantify cortical bone free water longitudinal relaxation time (30).

Here in this study, we used a Dual-TR STE-MR protocol with TE of 1.29 ms as a clinical tool for cortical bone free water quantification. A validation study on cadaveric bone samples *ex vivo* as well as a translational study on a cross-sectional population of thirty healthy subjects was performed.

## 2. Methods
### 2.1. Ex vivo study

A method for quantification of cortical bone free water was proposed using STE-MR imaging. The proposed strategy quantifies cortical bone free water concentration by comparing the mean signal intensity of cortical bone with that of a reference sample with known NMR parameters mimicking the cortical bone. Signal intensities were acquired by segmenting the cortical bone in Short Echo Time (STE) MR images. Cortical bone water was also quantified using dehydration (gravimetry) methods (21), to investigate the agreement between gravimetric measurement and STE-MRI measurements. To pursue this goal, cortical bone free water was measured in seven bovine cortical bone specimens, by two different methods and the agreement between the results was quantified using Bland and Altman analysis.

The whole procedure of the *ex vivo* experiments is summarized as a flow diagram in **Figure 1**.

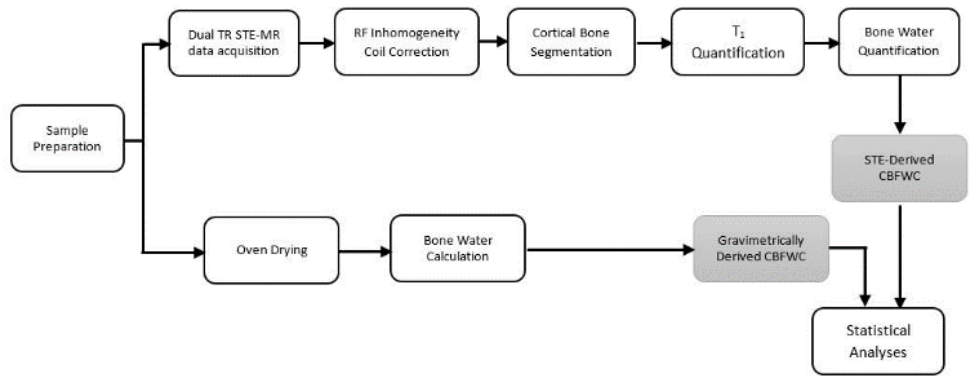

**Figure 1.** A simple flow diagram for the validation of cortical bone free water quantification technique using gravimetric experiment and STE-MR imaging. Cortical Bone Free Water Concentration (CBFWC) was measured using two different strategies, and their results were compared.

### 2.1.1. Sample preparation

Seven mature bovine tibial mid-shafts of newly slaughtered cows were bought from a local slaughterhouse and cleaned of muscle and soft tissue carefully. Seven plate-like specimens with an approximate dimension of $6.6 \times 15 \times 24$ mm$^3$ were cut out of each bone using an electric saw. All bovine cortical samples were cut from relatively the same location of the tibia, corresponding to the 38% of the tibia length as in in-vivo imaging. In order to make sure that the amount of dehydration that happens during the air drying is associated only to the bone (and not to any other type of tissue, like endosteum or periosteum), we ensured that the specimens were pure of cortical bone. To this end, the following three steps were iterated three times for each cut: i) immersing the specimen in an ultrasonic bath for 5 minutes (help the residual soft tissue get loosed), ii) hand grinding the specimen with silicon carbide paper, iii) cleaning the surface with an air compressor. The specimens were immersed in saline overnight to compensate for water loss that might have been occurred during specimen preparation. Then, the specimens were removed from saline, blotted dry and weighed to calculate the *wet weight*. After that, the specimens were stored in -20˚C till approximately 18 hours prior to MR imaging.

### 2.1.2. MR Imaging

All MR imaging was performed on a 1.5T MR scanner (Siemens, Magnetom Avanto) with an eight-channel receive knee coil. A *clinical* 3D Gradient Echo pulse sequence using Short Time of Echo (STE) on Gradients having maximum field strength of 45 mT/m and maximum slew rate of 200 mT/m/s was applied to obtain two series of axial images with two different repetition times (10 slices for each Repetition time (TR)) (previously published in (30, 31)).

Spins were excited by a full sinc Radio Frequency (RF) pulse with a duration of 2.5 ms. Imaging parameters were chosen to be: $TR_1/TR_2/TE = 20/60/1.29$ ms, field-of-view (FOV) = $267\times267$ mm$^2$, spatial resolution = $0.8\times0.8$ mm$^2$, slice thickness = 5 mm, flip angle = 20˚, readout Bandwidth = 781 Hz/Pix, number of slices = 10, total scan time of 20 minutes. Using the mentioned parameters, a stack of 20 STE-MR images was acquired out of which the cortical bone free water concentration (percentage) and cortical bone free water longitudinal relaxation time (ms) values were extracted through the steps described in the next sections.

We used 20% water in Deuterium Oxide (D$_2$O) doped with 27 mM MnCl$_2$ with longitudinal and transverse relaxation time parameters of 15 ms and 0.7 ms, respectively. Two vials were filled and adhered to the bone specimens in *ex-vivo* experiments and to the subjects' leg in *in-vivo* experiments.

### 2.1.3. Free water T$_1$ and T$_2^*$ Quantification

The proposed bone water quantification strategy demands $T_1$ and $T_2^*$ values of the target water molecules (free water). Previous studies have shown that cortical bone water $T_1$ depends on the subject and varies with age (30, 32). Therefore, T1 calculation must precede bone water quantification. The following steps computed free water $T_1$ values. 1) Segmenting the cortical bone in both images (short-TR/long-TR). 2) Computing the ratio value (r) as shown in Eq. 1 by dividing mean signal intensity of the segmented cortical bone acquired from the long-TR (TR2) image by that of the short-TR (TR1) image. 3) Calculating the cortical bone free water $T_1$-value for each slice by solving Eq. 1 using 'trust-region-dogleg' algorithm in MATLAB 7.14 (The MathWorks) (33). 4) Iterating the described three steps through all ten slices and computing ten T1-values per each subject. 5) Averaging all the obtained T1-values to be reported as the cortical bone free water longitudinal relaxation time.

$$r = \frac{1-\exp(-TR1/T1)}{1-f_z \exp(-TR1/T1)} \Big/ \frac{1-\exp(-TR2/T1)}{1-f_z \exp(-TR2/T1)} \quad (1)$$

$f_z$ in equation 1 is a function of $\tau/T_2^*$ (ratio of the pulse duration to $T_2^*$ relaxation of cortical bone) defining a correction parameter for relaxation ($T_2^*$) losses during RF excitation period (13).

For *ex vivo* study, since there was little variation in the age of the bovine specimens the value of free water $T_2^*$ was assumed to be 2.81 ms for the purpose of cortical bone free water quantification (19).

### 2.1.4. RF Coil Inhomogeneity Correction

As mentioned before, the quantification process was based on comparing the signal intensities of cortical bone and phantom. Hence even minor inhomogeneities of the transmit field or spatial dependence of the receive coil sensitivity may degrade the quantification process, largely. The transmit of the RF pulse was performed by the scanner's body coil which makes the transmit inhomogeneity of less concern. Experiments were performed with an eight-channel receive knee coil which demanded some correction for inhomogeneity in the received signal. By acquiring an image from a homogenous phantom (pure water) with the same imaging parameters, the inhomogeneity reception profile of the RF Coil was extracted. Thereafter, by simply pixel-wise dividing the STE-MR images of cortical bone by the STE-MR image of water phantom, the embedded RF Coil inhomogeneity was removed from the images.

### 2.1.5. Bone Water Quantification

Quantification of water concentration in cortical bone was performed by dividing the mean signal intensity of segmented cortical bone by that of an ROI placed on the phantom in the acquired STE-MR image using Equation (2):

$$\rho_{bone} = \rho_{ref} \cdot (SI_{bone} \cdot F_{ref})/(SI_{ref} \cdot F_{bone}) \cdot \exp(-TE\,(R_{2_{ref}}^* - R_{2_{bone}}^*)) \quad (2)$$

Where $\rho_{bone}$ and $\rho_{ref}$ are water concentrations of the bone and reference sample, and $SI_{bone}$ and $SI_{ref}$ are the mean signal intensities measured form STE-MR images, respectively. $R_{2_{bone}}^*$ and $R_{2_{ref}}^*$ are the effective transverse relaxation rates ($R_2^* = 1/T_2^*$) for bone and the reference sample respectively, and TE is the echo time. $F_{ref}$ and $F_{bone}$ are functions of relaxation times, pulse repetition time (TR), and the ratios $\tau/T_2^*{}_{bone}$ and $\tau/T_2^*{}_{ref}$.

By normalizing the longitudinal and transverse magnetization immediately after the end of the excitation and expressing them as $f_z(\tau/T_2^*)$ and $f_{xy}(\tau/T_2^*)$, respectively, F can be written as:

$$F = f_{xy}(\tau/T_2^*) \cdot (1 - \exp(-TR/T_1)) / (1 - f_z(\tau/T_2^*) \cdot \exp(-TR/T_1)) \quad (3)$$

As the pulse duration ($\tau$), is of the same order or longer than $T_2^*$, relaxation would occur during exciting spins and $f_z(\tau/T_2^*)$ and $f_{xy}(\tau/T_2^*)$ are the correction factors that corrects the steady-state signal for this purpose. We simulated Bloch equation distinctively for the phantom and cortical bone to calculate $f_{z_{ref}}$, $f_{xy_{ref}}$ and $f_{z_{bone}}$, $f_{xy_{bone}}$, separately.

### 2.1.6. Dehydration experiment

After performing MR measurements, the specimens were dehydrated in an oven at room temperature (21°C) for three days and were weighed to determine their "dry weight". Then, the percentage of water loss of the bone specimens (by weight) was calculated by dividing the difference between wet and dry weights by the dry weight. The calculated water content of cortical bone has been proven previously to be its free water content (21). Since the MR-derived free water concentration was in terms of volume percentage, the by-weight water loss obtained by the dehydration experiment was converted to by-volume water loss by simply dividing it by the specimen density.

## 2.2. In vivo study

We applied our method on a cross-sectional population of thirty healthy volunteers covering the age range of 20 to 70 years old (18F/12M) to quantify their cortical bone free water concentration and to monitor the variation of cortical bone free water concentration during aging.

### 2.2.1. Cortical Bone Segmentation

Cortical bone segmentation was highly critical in the *in vivo* quantification because accurate discriminations between marrow and bone in the periosteal boundary, and between connective tissues and bone in the endosteal boundary were quite challenging. Cortical bone pixels were selected manually by drawing polygons using Image J (National Institute of Health, US) software. For reducing errors caused by fallacious identification of cortical bone pixels as marrow/connective tissue or vice versa, the process of manual segmentation was repeated five times. The average of the mean signal intensities of the five repeated ROIs was computed and reported as the sole mean signal intensity of the whole cortical bone tissue in each slice. Figure 2 shows an example of an ROI placed on the cortical bone.

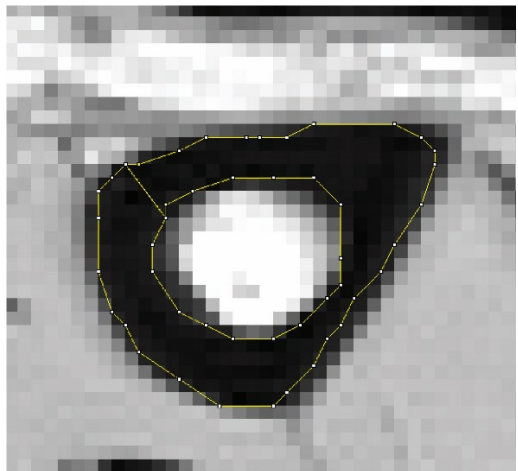

**Figure 2.** Segmented cortical bone extracted from STE-MR image of tibia, which was drawn manually using polygon tool in ImageJ software.

### 2.2.2. T1 and T2* quantification

$T_1$ values of cortical bone free water were quantified as it was quantified for *ex vivo* study (explained in ref 13). For this pilot study, in order to keep the total scan time in the limit of clinical scan times, we decided not to incorporate $T_2^*$ quantification in the protocol. Instead, a priori estimate of the free water $T_2^*$ was used for the purpose of quantifications. There is evidence in the literature supporting the fact that $T_2^*$ quantification might be pulse sequence-dependent (34, 35). Therefore, we used the STE pulse sequence to quantify free water $T_2^*$ of a small population of 6 healthy volunteers (3m/3f in the age range of 40 – 50 years).

We acquired five images of cortical bone using TE values of 1.29, 1.45, 1.6, 1.75, and 1.9 ms. As it is the basic idea of this project, these TE values are long enough to not capture the signal from bound water molecules. Then $T_2^*$ was quantified using a single component exponential fitting.

### 2.2.3. Subjects

The location of 38% of the tibia length, known to be the site of maximum cortical thickness (36), measured from the medial malleolus was selected to be the imaging site in the *in vivo* study. A cross-sectional population of 30 healthy subjects with a BMI < 30 kg/m$^2$ (12 males and 18 females), and in the age range of 20-70 years old was incorporated in the study. Half of the thirty patients have been previously reported (30). The prior article dealt with relaxometry of cortical bone free water concentration whereas in this study we developed a technique for quantification of cortical bone free water concentration and validated it in an *ex vivo* setup by using dehydration. Subjects who had medical histories that indicated disorders, surgery, or treatments (e.g. glucocorticoid therapy or antiepileptic drugs) compromising bone mineral homeostasis were excluded. The protocol was a clinical routine protocol and written informed consent was obtained from all subjects.

### 2.2.4. Reproducibility

Reproducibility, the ability of an entire experiment to be duplicated, is the measure of the precision of a method and it was examined for our proposed strategy by performing the whole procedure twice within the time distance of two months on each subject in a group of eight healthy volunteers (five males and three females). Root-mean-square difference between baseline and repeated measurements were calculated, as well as the intra-class correlation coefficient between the two measurements.

### 2.3. Statistical analyses

Statistical analyses were performed with MATLAB 7.14 (The Math Works) and Microsoft Excel. To evaluate the hypothesized agreement between STE-derived cortical bone free water and gravimetric measures in validation experiments, the Bland and Altman analysis was performed by using Microsoft Excel. For the *in vivo* studies, a regression line was calculated to show the association of age with cortical bone free water concentration and $T_1$. The root-mean-square difference and intra-class correlation coefficient were also calculated for measuring the reproducibility of the proposed strategy by using MATLAB.

## 3. Results

To demonstrate general image quality, Figure 3 shows STE-MR images both in *ex vivo* and *in vivo* setups. For the ex vivo images, the Signal-to-Noise-Ratio (SNR) values were reported to be 22 and 31 for TR = 20 ms and TR = 60 ms, respectively. Also, for the *in vivo* images, the SNR values were 18 and 28 for TR = 20 ms and 60 ms, respectively. Variations in signal intensity (shown in Figure 2) corroborates the fact that the applied MR protocol doesn't yield signal void for cortical bone.

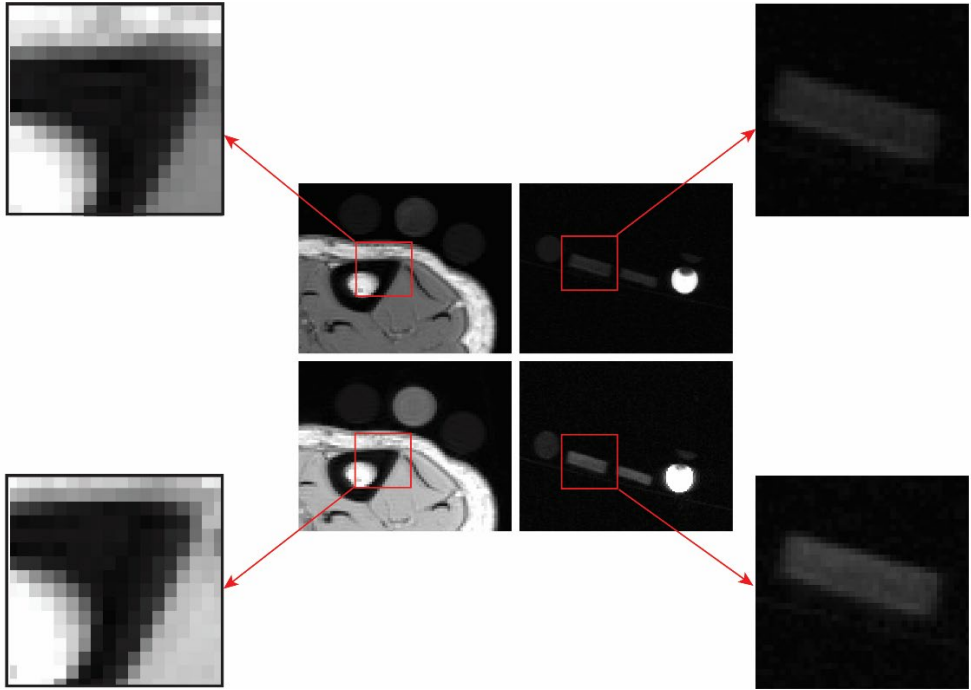

**Figure 3**. Shows the acquired STE-MR images of midshaft tibia along with reference samples in a 68-year-old male subject (right) and a bovine cortical bone specimen (left), for both TR = 20 ms (top) and TR = 60 ms (bottom. For each image a section of the cortical bone is zoomed in so that the variation of the signal intensities is easily noticed refuting the existence of signal void in the cortical bone area.

Cortical bone free water concentration values were calculated in seven bovine specimens using two different methods. The mean value of the cortical bone free water concentration obtained by STE-MR imaging using equation 1 assuming $T_2^*$ of free water as 2.81 ms (12) was reported as 4.37 $\pm$ 0.21 %, and mean value of cortical bone free water concentration obtained through dehydration method was reported as 5.34 $\pm$ 0.48 %. The mean value of cortical bone free water $T_1$ relaxation time for bovine specimens was also reported as 101.66 ms. Results are shown in detail in Table 1 for all samples.

**Table 1.** Calculated cortical bone free water concentration value of seven plate-like bovine cortical bone specimens by using two different methods

| Specimen # | Free water $T_1$ (ms) | Free water Concentration (%) by STE-MRI | Free water Concentration (%) by Dehydration |
|---|---|---|---|
| 1 | 60.1 | 4.48 | 5.69 |
| 2 | 108.91 | 3.99 | 4.51 |
| 3 | 99.2 | 4.45 | 5.55 |
| 4 | 97.81 | 4.71 | 6.12 |
| 5 | 123.27 | 4.31 | 4.95 |
| 6 | 101 | 4.36 | 5.48 |
| 7 | 121.35 | 4.32 | 5.13 |
| Mean ± SD | 101.66 ± 16.71 | 4.37 ± 0.21 | 5.34 ± 0.48 |

To investigate the agreement between the two methods (STE-MR and dehydration methods), we used both regression analysis and the analysis of differences (shown in Figure 4). Bland and Altman (B&A) plot describes the agreement between two quantitative methods by using the mean and standard deviation of the differences between the two measurements. The B&A plot, as well as the regression analysis, showed very good agreement between the two methods used for free water quantification.

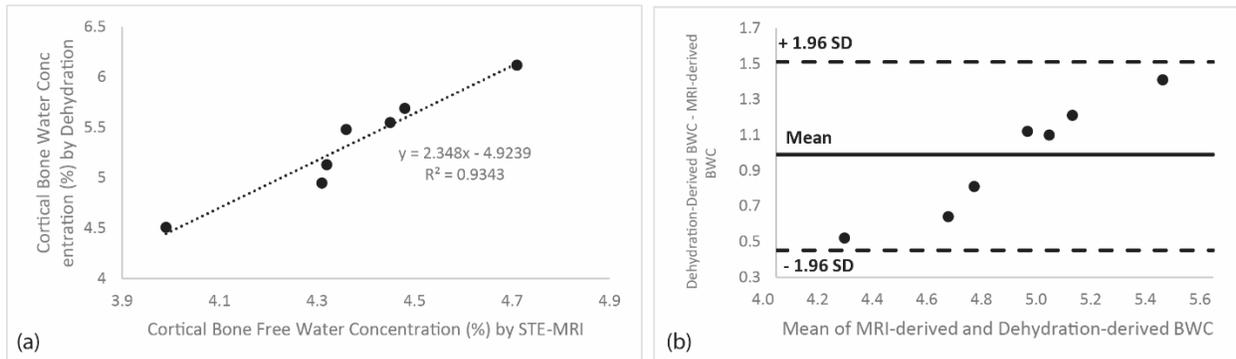

**Figure 4.** Free water concentration value is calculated using two different techniques for seven bovine cortical bone specimens. To assess the underlying agreement between two methods, correlation between them (a) and Bland-Altman plot with the representation of limits of agreement (b) are depicted, showing acceptable consistency between the two methods.

Since cortical bone free water concentration was hypothesized to be an indirect measure of cortical bone porosity and in the other hand porosity varies by aging, one can readily conclude that free water concentration may track the age-related variations of cortical bone porosity. Hence, the cortical bone free water concentration values in a cross-sectional population of thirty healthy volunteers covering the age range of 20-70 years old were quantified using the proposed methodology.

The $T_2^*$ values of cortical bone free water and the phantom were measured as 2.81 ± 0.31 ms and 0.7 ± 0.02 ms, respectively, through the mono-exponential fitting of signal intensities as a function of echo times. Using these values of $T_2^*$s, cortical bone free water was quantified among the subjects. The values of $T_1$ relaxation time ($T_{1free}$) and concentration ($\rho_{free}$) for cortical bone free water in thirty healthy volunteers were reported separately among five subsets of subjects in five different age decades. Mean and standard deviation of $T_1$ and water concentration values for each decade are shown in Table 2.

Table 2. Cortical bone free water concentration and T1 reported separately for 5 age subclasses.

| Age Decade | No. of subjects | Free water T1 (ms)* | Free water Concentration (%) * |
|---|---|---|---|
| 20-30 | 5: 4F/1M | 119.02 ± 16.34 | 2.41±0.11 |
| 30-40 | 6: 3F/3M | 156.47 ± 19.9 | 2.59 ± 0.15 |
| 40-50 | 8: 4F/4M | 164.86 ± 15.4 | 2.64 ± 0.19 |
| 50-60 | 5: 3F/2M | 196.74 ± 16.31 | 2.87 ± 0.21 |
| 60-70 | 4: 3F/1M | 221.75 ± 28.33 | 2.91 ± 0.06 |
| 70 + | 2: 1F/1M | 246.13 ± 13.86 | 3.29 ± 0.19 |
| Total | 30: 18F/12M | 173.86 ± 40.95 | 2.71 ± 0.28 |

*Data are means ± Standard Deviations.

The average of cortical bone free water $T_1$-values and cortical bone free water concentration for all thirty healthy volunteers was computed as 173.86 ±40.95 ms and 2.71% ± 0.28, respectively.

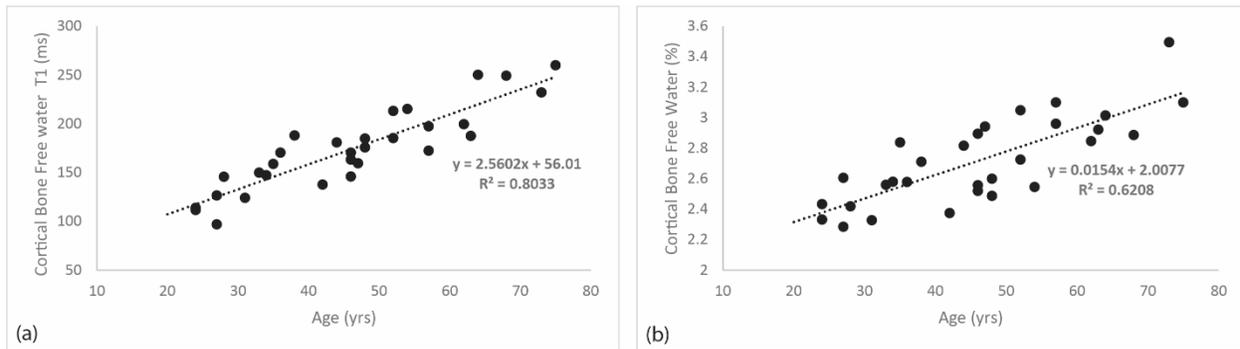

Figure 5. The association with age was examined for cortical bone free water concentration ($\rho$free) and free water T1 in a cross-sectional population of free water concentration and T1free in thirty healthy volunteers covering the age range of 20-75 years old.

The association of cortical bone free water concentration ($\rho_{\text{free}}$) and cortical bone free water longitudinal relaxation time ($T_{1\text{free}}$) with age were depicted in figure 5. As the Pearson correlation coefficients testified ($r^2 = 0.62$ for $\rho_{\text{free}}$, $r^2 = 0.8$ for $T_{1\text{free}}$), bone water concentration and cortical bone free water $T_1$ are possible surrogate measures of porosity as previously were demonstrated by the authors in several other studies (22, 30).

Evaluation of reproducibility of our proposed strategy for cortical bone free water quantification was essential in our future longitudinal studies and therefore eight healthy subjects have undergone this examination twice within two months. Figure 6 shows the results of reproducibility a scatter plot of test-retest data. The results indicate reproducibility with an intra-class correlation coefficient of 0.95.

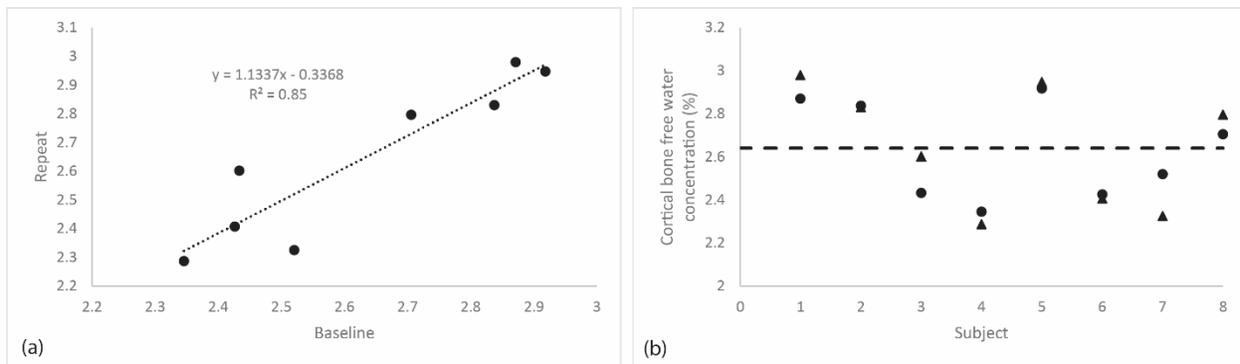

**Figure 6.** Reproducibility data depicted in scatter plot (a) and test-retest format (b). Data indicate intra-class correlation coefficient (ICC) of 0.95. Analysis of variance suggests significant difference among subjects ($p < 0.0001$).

## 4. Discussion

Our work used dual-TR STE-MR imaging as a clinical MR solution to quantify cortical bone free water concentration without contamination of bound water. The validity of the method was investigated through an *ex vivo* experiment on seven plate-like bovine cortical bone specimens. The chosen TE value in our study guaranteed the fact that the detected signal majorly was emanated from the cortical bone free water pool. Cortical bone bound water $T_2^*$ was reported in the literature at 1.5T to be 0.45 ms. Therefore, our employed TE (1.29 ms) resulted in a 95% loss of bound water signal indicating that our proposed strategy filtered bound water signal out. The clinical motivation beyond quantifying cortical bone free water was the fact that its parameters are key factors for both assessing bone quality and diagnosis of bone-related diseases (37-39).

Before testing our method in clinical practice for thirty healthy volunteers, we investigated whether the measurements made by the dehydration technique are reproducible by our new method or not. Since the two measurements were made on the same scale, we quantified their agreement by Bland and Altman analysis. According to this analysis, there was a little suggestion of systematic bias between the methods. Thus, STE-MR-based method tends to give a lower value, by 0.99%. Despite this, the limits of agreement (0.04 and 0.7) are small enough for us to be confident that the new method can be as efficient as the established one for cortical bone free water quantification.

The systematic bias between the methods suggests that dehydration water concentration is significantly higher than STE-MR derived water concentration, which was also previously shown by Chen *et al*. (40) and Biswas et al. (21). This discrepancy stems from the fact that 3-day air-drying removes loosely bound water plus free water from all pores of cortical bone ranging from small space of lacunae to the large space of giant canals. While, STE-MRI, on the other hand, captures signal only from the water molecules residing in giant canals of cortical bone that have long T2 values.

In a similar study, Du *et al*. quantified cortical bone free water concentration of bovine specimens by using two different methods; bi-component analysis of UTE-MR images and dehydration (21). The mean value of cortical bone free water concentration for fourteen specimens, which was extracted from four bovine tibiae, was reported as about 2.88% and 5.77% for UTE-assessed and 3-day air drying, respectively. Our reported data for gravimetrically-derived free water concentration as 5.37% was in a good agreement with their study. STE-MRI derived free water concentration reported in our study was higher than their UTE-MRI derived which is opposite to what we expected. This might be due to the underestimation of cortical bone free water signal fraction using bi-component analysis of UTE as inhomogeneous line broadening of $T_2^*$ might affect the process of discrimination (23).

Cortical bone free water $T_1$ was previously quantified by different groups as follows: Horch *et al.* reported $T_{1free}$ to be in the range of 500-1000 ms at 4.7T. They used inversion recovery preparation pulses with twenty-four different recovery times preceding a Carr-Purcell-Meiboom-Gill to acquire a T1-T2 spectrum. Given the corresponding $T_2$ of the free water from the previously acquired $T_2$ spectrum, they measured $T_1$ value (7). In another study performed previously by the authors, cortical bone free water $T_1$ was quantified by using 3D hybrid radial ultrashort MR imaging (3DHRUTE) followed by a model-based post-processing optimization technique at 3T. The average cortical bone free water $T_1$ was reported to be 306 ms for forty healthy volunteers (22). Regarding the fact that the longitudinal relaxation time of the tissue reduces by reducing the main magnetic field (30, 41), the quantified value of $T_{1free}$ was expected to be less than the reported values at 4.7T and 3T showing the decreasing trend. Therefore, the $T_1$ values for cortical bone free water reported in our study were in good agreement with the literature.

Different researchers also quantified cortical bone free water concentration previously. Manhard *et al.* (27) applied double adiabatic full-passage on a population of five healthy volunteers to quantify the absolute value of free water concentration and reported the mean value to be 7.32 moles of hydrogen ($^1H$) per liter of bone. In the other study performed by the authors (22) cortical bone bulk water concentration was quantified for a population of forty healthy volunteers by using 3DHRUTE pulse sequence and then a model-based optimization algorithm was used to calculate free water concentration out of the bulk concentration. The mean value was reported as 5.89% which is consistent with the reported value in this study. Chen et al conducted a bi-component analysis of UTE-MRI signal to quantify bound and free water. Their study was performed on 13 human cortical bone cadaveric samples and cortical bone free water concentration was reported as 4.7% - 5.3% (40). All of these free water quantification strategies were involved with advanced sequence design using ultrashort echo times. Manhard *et al* used adiabatic RF pulses to prepare (invert or saturate) the spins and the other two mentioned methods used Radial sampling to fill most of the k-space before the decay of the signal in UTE sequences. As opposed to them, our proposed method used a product sequence of Siemens MR scanner which is readily available in the clinic for everyone.

High correlation coefficients were reported between cortical bone free water parameters (concentration ($r^2 = 0.62$) and longitudinal relaxation time ($r^2 = 0.8$) values) and age. During aging, as the consequence of the enlargement of cortical pores, their surface-to-volume ratio (S/V)) decreases and it has two consequences; the concentration of cortical bone free water, as well as the mobility of free water molecules, increase. The former justifies the high correlation between $\rho_{free}$ and age while the latter justifies the high correlation between $T_{1free}$ and age. By increasing the mobility of the free water molecules, the mechanism of energy transfer between the lattice and the spins is hindered and therefore the $T_{1free}$ values increase.

The correlation between free water concentration and age is relatively lower than that of free water $T_1$. It might be due to a limitation of our work which is related to $T_2^*$ quantification. Cortical pores have different geometry for different individuals that also changes during aging and will affect the bone-air interface and consequently affect the susceptibility of free water in cortical bone. Therefore, similar to $T_1$, free water $T_2^*$ also depends on the subject. However, in this pilot study in order to keep the scan time in the clinical range, we used a priori estimation of cortical bone free water $T_2^*$ by quantifying and averaging its value in a small population of subjects.

The significant correlation between STE-MR-derived free water concentration and age shows that it could be a possible surrogate measure of porosity that increases by ageing. Prior to our study, some research groups have also tried to introduce a biomarker capable of predicting the age-related increasing trend of porosity. Rajapakse *et al.* acquired UTE images with multiple echo times and introduced the ratio of the signal intensities of the shortest possible TE to the longest echo as the porosity index (PI). They reported a significant correlation ($r^2 = 0.64$) between PI and age among sixteen (9F/7M) cadaveric human cortical bone specimens covering the age range of 37 to 93 years old. (42).

Li *et al.* introduced Suppression Ratio (SR) as the ratio of unsuppressed UTE signal intensity to the long-T2-suppressed signal intensity being a potential biomarker of porosity. They quantified SR for a cross-sectional population of 72 healthy subjects (20 – 80 years) and reported a correlation coefficient of $r^2 \cong 0.5$ with age. This correlation suggested that SR can be considered as a potential measure of porosity but with one limitation; it fails to fully differentiate between free and bound water (43).

The same potential was investigated for our proposed biomarkers ($T_{1free}$ and $\rho_{free}$) by quantifying them for a cross-sectional population of thirty healthy volunteers. The obtained correlation coefficient ($r^2 = 0.62$) suggested that STE-MR-derived water concentration is as efficient as porosity index (PI) and more efficient than suppression ration (SR) for modeling the age-related increase in cortical porosity.

There are several limitations to our study as follows: firstly, $T_1$ quantification was performed using only two measurements, which might introduce errors in the quantified values. Although the values were in good agreement with the literature, both in terms of its decreasing trend by increasing $B_0$ and in terms of its association with age, there is still room for improving our $T_1$ quantification strategy in future studies.

Secondly, there are protons related to the lipid methylene that are long-T2 components and might contaminate the STE signal. This fat contamination could be a source of signal cancellation between free water and fat. A bi-component analysis of STE-MRI signal is in order for the future research of our group.

Thirdly, there was limitation related to the subject's motion in the *in vivo* experiments. Despite all efforts to immobilize subjects' legs during imaging, unwanted motions might have occurred, which incurs inaccuracy in image analysis and quantification procedure.

Fourthly, cortical bone free water parameters can be quantified through two different strategies: Voxel-based and ROI-based quantification. Although voxel-based quantification was more appropriate, errors attributed to misregistration between two images ($TR_1$ and $TR_2$) and minor inconsistencies in segmentation degraded its accuracy. Therefore, we chose ROI-based quantification as a more reliable approach by which we neglected the spatial distribution of $T_1$ and $\rho$ and considered the cortical bone tissue as a homogenous medium implicitly.

Since STE-MR imaging is available in all clinical environments, the proposed method benefits from clinical compatibility which is of great importance. It also obviated the challenge of the conflation of free and bound water since it was shown to be sensitive to only free water molecules of cortical bone.

## 5. Conclusion

Cortical bone free water longitudinal relaxation time ($T_{1\text{free}}$) and cortical bone free water concentration ($\rho_{\text{free}}$) was quantified by clinically available dual-TR STE-MR imaging. This protocol has the potential for clinical use and complementary role to the UTE-based techniques for cortical bone water quantification. These parameters monitored age-related alterations of cortical bone in a population of thirty healthy volunteers.

## 6. Acknowledgement

All the imaging was done in Payambaran Hospital and hereby the authors wish to thank Mr. Mohsen Shojaee-Moghaddam and Mrs. Niloofar Tondro for their continuous help and support during imaging. We also wish to thank Mr. Masood Yarmahmoodi for providing us with all the equipments required for specimen preparation. The authors would like also to acknowledge the comments of the anonymous reviewers that contributed to the improvement of this work.